%% file: main.tex
\documentclass[aps,prl,superscriptaddress,showpacs,twocolumn,floatfix]{revtex4}

\usepackage{graphicx}
\usepackage{floatflt}
\usepackage{tabularx}

\def\gtorder{\mathrel{\raise.3ex\hbox{$>$}\mkern-14mu
 \lower0.6ex\hbox{$\sim$}}}
\def\ltorder{\mathrel{\raise.3ex\hbox{$<$}\mkern-14mu
 \lower0.6ex\hbox{$\sim$}}}

\begin{document}

\title{A detailed study of the nuclear dependence of the EMC effect and short-range correlations}

\author{J.~Arrington}
\affiliation{Physics Division, Argonne National Laboratory, Argonne, IL, 60439, USA}

\author{A.~Daniel}
\affiliation{Ohio University, Athens, OH, 45701, USA}
\affiliation{University of Virginia, Charlottesville, VA, 22904, USA}

\author{D.~B.~Day}
\affiliation{University of Virginia, Charlottesville, VA, 22904, USA}

\author{N.~Fomin}
\affiliation{Los Alamos National Laboratory, Los Alamos, NM, 87545, USA}

\author{D.~Gaskell}
\affiliation{Thomas Jefferson National Accelerator Facility, Newport News, VA, 23606,  USA}

\author{P.~Solvignon}
\affiliation{Thomas Jefferson National Accelerator Facility, Newport News, VA, 23606,  USA}

\date{\today}

\begin{abstract}
\begin{description}

 \item[Background:] 
The density of the nucleus has been important in explaining the nuclear dependence of the quark distributions, also known as the EMC effect, as well as the presence of high-momentum
nucleons arising from short-range correlations (SRCs).  Recent measurements
of both of these effects on light nuclei have shown a clear deviation from
simple density-dependent models.

\item[Purpose:]    
 A better understanding of the nuclear quark distributions
and short-range correlations requires a careful examination of the experimental data on these effects to constrain models that attempt to describe these phenomena.

\item[Methods:]   
We present a detailed analysis of the nuclear dependence of the
EMC effect and the contribution of SRCs in nuclei, comparing to predictions 
and simple scaling models based on different pictures of the underlying
physics. We also make a direct, quantitative comparison of the two effects to
further examine the connection between these two observables related to nuclear
structure.  

\item[Results:]    
 We find that, with the inclusion of the new data on light nuclei,
neither of these observables
can be well explained by common assumptions for the nuclear dependence.
The anomalous behavior of both effects in light nuclei is consistent
with the idea that the EMC effect is driven by either the presence of high-density configurations
in nuclei or the large virtuality of the high-momentum nucleons associated with these configurations.
\item[Conclusions:] 
The unexpected nuclear dependence in the measurements of the
EMC effect and SRC contributions appear to suggest that the local
environment of the struck nucleon is the most relevant quantity for explaining
these results.  The common behavior suggests a connection between the two
seemingly disparate phenomena, but the data do not yet allow for a clear
preference between models which aim to explain this connection.
\end{description}

\end{abstract}
\maketitle


\input{intro}

\input{adep_emc}

\input{adep_src}

\input{src_vs_emc}

\input{results}

\input{summary}


\begin{acknowledgments}
We thank Wally Melnitchouk, Doug Higinbotham, Larry Weinstein, Don Geesaman,
Or Hen, Eli Piasetzky, Maarten Vanhalst, Jan Ryckebusch, and Wim Cosyn for
fruitful discussions.  We also thank Sergey Kulagin for providing calculations
of separation energies used in our analysis.
This work supported by the U.S. DOE through contracts
DE-AC02-06CH11357, DE-AC05-06OR23177, and DE-FG02-96ER40950 and research grant
PHY-0653454 from the National Science Foundation.
\end{acknowledgments}

\bibliography{main}

\end{document}

%% file: intro.tex
\section{Introduction}

The nucleus is a system of strongly-interacting protons and neutrons. The
characteristic scale for the nucleon momentum is the Fermi momentum, $k_F
\approx200$--270~MeV/c, a consequence of the interaction of the nucleon with
the mean field of the nucleus. The strongly repulsive nature of the
nucleon-nucleon (NN) interaction at short distances prevents two nucleons from
coming very close together and this loss of configuration space demands
the existence of high-momentum components in the nuclear ground state wave
function. These can not be described in the context of mean field models and
are commonly called short-range correlations (SRCs). Inelastic electron
scattering was suggested long ago~\cite{Czyz196347} to be a source of
qualitative information on SRCs, yet they remain one of the least-well
characterized aspects of the structure of stable nuclei.

Knockout reactions studied in inclusive and exclusive electron
scattering~\cite{frankfurt93, arrington99, arrington01, egiyan03, egiyan06,
shneor07, subedi08, fomin2012} have isolated SRCs by probing the high-momentum
tail of the nuclear momentum distribution.  The tail is assumed to be the
result of short-range hard interactions between nucleons~\cite{frankfurt93,
sargsian03, arrington11}, allowing a study of short-distance structure via
reactions with high-momentum nucleons.  The strength of SRCs in the nucleus
has long been assumed to scale with nuclear density~\cite{Antonov86,
frankfurt93, egiyan03, sargsian03, egiyan06}, a proxy for the probability of
two nucleons interacting at short distances.

Typical parametrizations of the repulsive core of the NN
interaction~\cite{Lacombe:1980, wiringa95, machleidt01} show a sharp rise in
the potential well below 1~fm.  Because the nucleon has an RMS radius of
roughly 0.85~fm~\cite{zhan11}, nucleon wave-functions can have significant overlap.  In
heavy nuclei, the typical inter-nucleon separation is 1.6~fm, suggesting that
the nucleons have some overlap most of the time, and this short-range
interaction may cause a modification of the structure of the nucleon.  There
is a long history of searches for this kind of ``medium modification'' of
nucleon structure through measurements of the in-medium nucleon form
factors~\cite{VanDerSteenhoven:1986vd, strauch02, Paolone:2010qc} or
modification of the quasielastic response in nuclei~\cite{Sick:1986pt,
mckeown86, Jourdan:1996ut, Morgenstern:2001jt, Carlson:2001mp}. Overlap of the
nucleon wave-functions may also allow for direct quark exchange, providing a new mechanism
for modifying quark momentum distributions in the nucleus and one may expect
them, like SRCs, to have a dependence on the average nuclear density.

The modification of the quark momentum distributions was first observed by the EMC collaboration~\cite{aubert83} and is
commonly referred to as the EMC effect.  It was discovered that the per
nucleon cross section in deep inelastic scattering (DIS) was different for iron and the
deuteron.  Because the binding energy of nuclei is extremely small compared to
the energy scales in DIS, the early assumption was that the parton distribution functions
(pdfs) of the nucleus would be a simple sum of the proton and neutron pdfs,
except at the largest values of the quark momentum fraction (Bjorken-$x$)
where the Fermi motion of the nucleus becomes important.  Since the DIS cross
sections depend on the quark distributions, the difference in the measured
cross sections for iron and the deuteron indicated a suppression of quark pdfs
in nuclei for $0.3 < x < 0.7$, and the size of this effect was seen to scale
with the nuclear density.

Thus, the nuclear density has often been taken as the underlying cause
of both the $A$ dependence of the nuclear pdfs and the presence of
short-distance configurations which give rise to high-momentum
nucleons.  Because of this, it is natural to assume that the behavior of both
the EMC effect and the presence of SRCs will be closely connected.  The
relationship between these two effects was recently
quantified~\cite{weinstein2010rt}, via a linear correlation between the SRCs
in the tail of the nucleon momentum distribution and the size of the EMC
effect.

While the EMC measurements performed in the '80s and '90s were well
described by a density-dependent fit~\cite{gomez94}, the weak $A$
dependence for these nuclei could be equally well described in other
approaches that have been proposed~\cite{geesaman95, norton03}.
For example, some works have explained the effect in terms of the average
virtuality ($\nu=p^2-m_N^2$) of the nucleons~\cite{sargsian03, gross92, ciofi2007}, connecting it more
closely to the momentum distributions. Given the limited precision of the
EMC effect measurements and the fact that it
grows smoothly but slowly for heavy nuclei, it is difficult to make a clear
determination of which approach best describes its $A$ dependence.

Recent measurements on light nuclei~\cite{seely09, fomin2012} have
observed a clear breakdown of the density-dependent picture for both the
nuclear modification of quark pdfs and the strength of short-range
correlations in nuclei, while still preserving the linear correlation between
the two~\cite{Hen:2012fm}.  In this work, we provide a detailed analysis of
the nuclear dependence of these two quantities, focusing on comparisons to
model-inspired assumptions.  We also perform an extended version of the analysis
presented in Refs.~\cite{weinstein2010rt,Hen:2012fm}, aimed at testing the
possible explanations for the correlation. For both the
analysis of the $A$ dependence and the direct comparison of the EMC and SRC
data, we examine in more detail the meaning of the observables associated with
these effects.  As the underlying dynamics behind the examination of the
direct correlation differ, additional corrections may be required when
comparing the observables that are typically associated with the EMC effect or
the presence of SRCs.

%% file: adep_emc.tex
\section{Nuclear Dependence of the EMC Effect}
\label{adep_emc}

Deep inelastic scattering provides access to the quark distributions in nuclei via measurements of inclusive cross
sections. This cross section for electron or muon scattering
from a nucleus can be written as
\begin{equation}
\frac{d\sigma}{dx dQ^2} = \frac{4\pi \alpha^2 E'^2}{x Q^4}\frac{E'}{E} 
\left[ F_2 \cos^2{\frac{\theta}{2}} 
+  \frac{2\nu}{M} F_1  \sin^2{\frac{\theta}{2}} \right],
\end{equation}
where $F_1$ and $F_2$ depend on $x$ and $Q^2$.
In the parton model, information about the quark distribution
functions is encoded in the $F_1$ and $F_2$ structure functions.
In the Bjorken limit ($Q^2$, $\nu \rightarrow \infty$, fixed
$\frac{\nu}{Q^2}$), the structure functions become independent of $Q^2$,
\begin{equation}
F_1(x) = \frac{1}{2} \sum_q e^2_q q(x),~~~ F_2(x) = 2 x F_1, 
\end{equation}
where $q(x)$ is the quark distribution function and $e_q$ is the quark
charge for a given flavor ($u$, $d$, $s$).

The per-nucleon ratio of the $F_2$ structure functions between 
an isoscalar nucleus and the deuteron is then a direct measure of the modification
of quark distributions in nuclei. Experimentally, this ratio is
defined as $R_{EMC} = (F^A_2/A)/(F^D_2/2)$. The deuteron structure function
in the denominator is taken to approximate the sum of free proton and neutron
structure functions. In almost all measurements of the EMC effect, an 
additional assumption is made that the ratio of longitudinal to transverse cross 
sections, $R=\sigma_L/\sigma_T$, is $A$-independent such that 
the unseparated ratio of cross sections corresponds directly to the $F_2$ ratio, 
i.e., $\sigma_A/\sigma_D = F^A_2/F^D_2$.  For non-isoscalar nuclei an additional correction is typically applied to account for the difference in DIS cross sections between protons 
and neutrons. 

\begin{figure}[htb]
\centering
\includegraphics[angle=270,width=76mm]{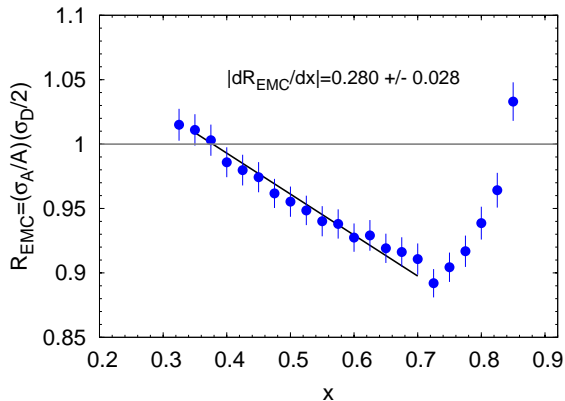}
\caption{(color online) EMC ratio, $(\sigma_A/A)/(\sigma_D/2)$, for
  carbon~\cite{seely09}. Solid line is a linear fit for 0.35$<x<$0.7. }
\label{fig:emc_example}
\end{figure}

Figure~\ref{fig:emc_example} shows a measurement of the EMC ratio for
carbon from Ref.~\cite{seely09}. The region from $x=0.3$ to $0.7$ shows
the depletion in the cross section ratio characteristic of all nuclei. The 
increase of the cross section ratio at large $x$ is attributed
to the greater Fermi momentum in the heavy nucleus
as compared to the deuteron. The shape of the EMC ratio
appears to be universal, independent of nucleus, while the magnitude
of the suppression generally increases with $A$.

The origin of the EMC effect has been a topic of intense theoretical discussion
since its original observation. There have been many explanations
proposed, and these can be broadly broken down into two categories.
The first includes only ``traditional'' nuclear physics effects,
using convolution models with binding effects, detailed models
of the nucleon momentum distribution, or pion-exchange contributions.
The other category invokes more exotic explanations such as re-scaling of
quark distributions in the nuclear environment, contributions of six or nine
quark bags, or modification of the internal structure of the nucleons
such as ``nucleon swelling'' or suppression of point-like nucleon
configurations. Several reviews give an overview of models of the EMC
effect~\cite{geesaman95, norton03, piller99}.

Because the observed suppression of the $F_2$ structure function between
0.3$<x<$0.7 is relatively straightforward to reproduce in a variety of
approaches, it is difficult to evaluate the different models without also
carefully examining the $A$ dependence. Direct comparisons to calculations
are limited by the fact that many calculations are made for nuclear matter,
and extrapolated to finite nuclei by assuming some simple scaling with $A$ or nuclear
density.  Others use more realistic nuclear structure
input~\cite{CiofidegliAtti:1991ae, kulagin2006, cloet06}, but often do not
include light nuclei.  So rather than comparing directly to calculations of
the EMC effect, we will examine its nuclear dependence,
comparing the data to different parameters assumed to drive the modification
of the nuclear pdfs.

In our examination of the $A$ dependence, we use the
data from SLAC E139~\cite{gomez94} and the recent data on light nuclei
from Jefferson Lab E03-103~\cite{seely09}. SLAC E139 sampled a range of nuclei
from  $A=4$ to $197$, allowing a large lever arm for studying the nuclear
dependence. Jefferson Lab experiment E03-103 adds $^3$He and
additional precise data on $^4$He, $^9$Be, and $^{12}$C.  While the JLab data
is at somewhat lower $Q^2$ values than the SLAC data, it has been
shown that the target ratios in this $Q^2$ range have very little deviation
from the DIS limit, even for $W^2$ values below 4~GeV$^2$~\cite{arrington06a,
seely09, Daniel:2007zz}.  The E03-103 data that go into extracting the EMC
slope examined in this work are all very near or above $W^2$=4~GeV$^2$.

We use the definition of the ``size'' of the EMC effect as introduced
in~\cite{seely09}, i.e., $|dR_{EMC}/dx|$, the value of the slope of a linear
fit to the cross-section ratio for $0.35<x<0.7$.  These limits were chosen to
give a range of high precision data whose behavior was linear, but the
extracted slope is not very sensitive to small changes in the $x$ region
chosen. This definition reduces the sensitivity to normalization errors, which
would otherwise be significant if one were to assess the nuclear dependence at
a fixed value of $x$, especially for light nuclei.  The impact of
normalization uncertainties for the deuteron measurements (common to all
ratios in a given experiment) is also reduced in this approach.  This
procedure makes use of the fact that the EMC effect has a universal shape for
$x>0.3$, exhibited by all experimental data.

Table~\ref{tab:2n_test} lists the EMC slopes extracted from the two
data sets described above. We do
not include data from earlier measurements due to their relatively
poor precision and/or limited $x$-coverage.

\begin{table}[htb]
\begin{center}
\caption{Combined EMC results from JLab E03-103~\cite{seely09} and
SLAC E139~\cite{gomez94} (averaged over $Q^2$). For JLab data,
$|dR_{EMC}/dx|$ was extracted in the
$0.35\le x \le 0.7$ range. SLAC data, whose binning was different,
were fit over $0.36\le x \le 0.68$.  For both cases, statistical and
point-to-point systematic uncertainties were applied to each $x$-bin
and the normalization uncertainties (including the 1\% normalization
uncertainty on deuterium common to all ratios for the SLAC data) were
applied to the extracted slope.
}
\label{tab:2n_test}
\begin{tabular}{|c|c|c|c|}
\hline
A & JLab & SLAC &Combined\\
\hline
$^3$He  & 0.070$\pm$0.028 & --              &0.070$\pm$0.028 \\
$^4$He  & 0.198$\pm$0.027 &0.191$\pm$0.061  &0.197$\pm$0.025 \\
Be      & 0.271$\pm$0.030 &0.208$\pm$0.038  &0.247$\pm$0.023 \\
C       & 0.280$\pm$0.029 &0.318$\pm$0.041  &0.292$\pm$0.023 \\
Al      &        --       &0.325$\pm$0.034  &0.325$\pm$0.034 \\
$^{40}$Ca &--             &0.350$\pm$0.047  &0.350$\pm$0.047 \\
Fe      &--               &0.388$\pm$0.032  &0.388$\pm$0.033 \\
Ag      &--               &0.496$\pm$0.051  &0.496$\pm$0.052 \\
Au      &--               &0.409$\pm$0.039  &0.409$\pm$0.040 \\
\hline
\end{tabular}
\end{center}
\end{table}

When considering the nuclear dependence of the EMC effect, it is
important to be aware of corrections which depend on $A$ or
$Z$, such as Coulomb distortion~\cite{aste05}.
The influence of the Coulomb field of the nucleus on the incident or
scattered lepton is a higher order QED effect, but is not typically
included in the radiative corrections procedures.   In addition, the size of EMC effect is taken directly from the cross section 
ratio instead of the structure function ratio, thus assuming no nuclear dependence 
in $R=\frac{\sigma_L}{\sigma_T}$. Coulomb distortions introduce kinematic
corrections and consequently have a direct effect on the extraction of $R$. 
An indication of nuclear dependence in $R$ was observed 
recently~\cite{Solvignon:2009it} after applying Coulomb corrections to SLAC 
E139 and E140~\cite{dasu94} data. 
Coulomb distortion was accounted for in the JLab data, but not the
SLAC data, where it is
estimated to be negligible for nuclei lighter than $^{12}$C and at most a 2\% effect on the 
$^{197}$Au EMC slope.  These changes do not significantly affect the
nuclear dependencies studied below.

The JLab and SLAC data also used different
prescriptions to correct non-isoscalar nuclei.  In the case of SLAC data, a
simple, $x$-dependent parametrization was employed based on high $Q^2$
data for $F_2^D/F_2^p$.  A more sophisticated correction was applied
to the JLab data~\cite{seely09}, using a smeared ratio of free proton and
neutron cross sections~\cite{Arrington:2008zh}.  Reanalysis of the SLAC data
using the updated isoscalar corrections yields slightly larger EMC slopes for
the very heavy nuclei, but does not impact the overall conclusions of this analysis.
A detailed comparison of these effects for both the SLAC data and the
heavy target data from JLab E03-103 is in progress~\cite{Daniel:2012, Solvignon:2012}.

Early calculations of the EMC effect included only the impact of Fermi
motion and were
unable to give a significant suppression at large $x$.  One can go beyond
simple smearing by including the effect of the binding energy of the nucleus.
However, the impact of the average nuclear binding is small and peaks $A=56$,
while the EMC effect continues to grow in heavier nuclei.  Thus, the binding
energy per nucleon, $E_A/A$, cannot explain the full modification of the
nuclear pdfs~\cite{Miller:2001tg}.  

\begin{figure}[htb]
\centering
\includegraphics[angle=270, width=78mm]{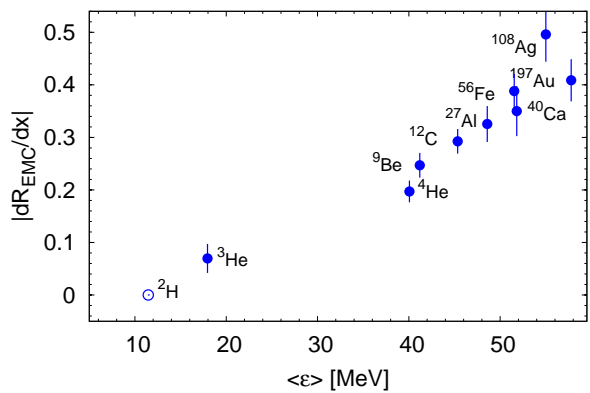}
\caption{(color online) Magnitude of the EMC effect, 
$|dR_{EMC}/dx|$, vs. the average nucleon separation energy. The empty circle
indicates the known (zero) deuteron EMC slope.}
\label{fig:emc_esep}
\end{figure}

While the nuclear binding energy is insufficient to explain the EMC effect,
high-energy electron scattering involves a near instantaneous scattering
and averages over all nucleons in the target.  This suggests that the
average separation energy may be the more relevant quantity in evaluating
the effect of binding.
The heart of the binding model describes nucleons bound in a nucleus with some
non--zero three-momentum, and as a consequence of the nuclear binding, an
energy modified from its usual on--shell value, i.e., $E_N \ne \sqrt{p_N^2 +
m_N^2}$. The bound nucleon has a removal or separation energy $\epsilon$, with
its total energy given by $E_N = m_N + \epsilon$ (ignoring the kinetic energy
of the recoiling nucleus). In practice the average separation energy is often
determined using the Koltun sum rule~\cite{koltun_sr},
\begin{equation}
 \langle \epsilon \rangle + \frac{\langle p^2 \rangle}{2 m_N} = 2 \frac{E_A}{A},
 \label{eq:koltun}
\end{equation}
where $p$ is the nucleon three-momentum and $E_A/A$ is the binding energy per
nucleon. An alternate formulation of the above rule was by proposed in
Ref.~\cite{dieperink74}, incorporating a recoil factor, but is
not used in the analysis presented here. The modification of the nucleon energy results in a value of $x=Q^2/2
p_N \cdot q$ shifted by $\approx \langle \epsilon \rangle /m_N$. 
In this context, the depletion of the cross section for $A>$2 in DIS
is associated with off-shell nucleons and binding produces a
simple rescaling of the relevant kinematic variable ($x$) and does
not imply an inherent modification of the nucleon structure in the nucleus.
Refs.~\cite{bickerstaff89,geesaman95} gives excellent overviews of early
calculations of the EMC effect in the binding approach. This approach was
relatively successful in reproducing the shape of the EMC effect at large
$x$~\cite{li88, CiofidegliAtti:1991ae, kulagin2006}, although calculations
including only this effect consistently underpredict the observed EMC effect.

Figure~\ref{fig:emc_esep} shows the extracted EMC ratio as a function of the
average nucleon separation energy, $\langle \epsilon \rangle$
from~\cite{kulagin_priv}, which provides the most complete set of
nuclei. In this figure, the separation energy was calculated
from spectral functions used and described in~\cite{kulagin2006, kulagin2010};
they include contributions from both mean--field and correlated (high-momentum) components of the nuclear wave function. While the separation energy
is an inherently model--dependent quantity, we have investigated alternate
definitions of the separation energy based on the Koltun sum rule as given
in Eq.~\ref{eq:koltun} and found that the typical agreement is usually better 
than 5 MeV.  However, some calculations use modified estimates of $\langle
\epsilon \rangle$, which can yield larger disagreements.

Qualitatively, the size of the
EMC effect correlates very well with the average separation energy, as was
also observed in another recent analysis~\cite{Benhar:2012nj}, using a slightly
different measure of the EMC effect and modified calculation of the mean
separation energy. However, while the correlation with the EMC effect is
good, detailed calculations based on the binding associated with the mean
separation energy~\cite{CiofidegliAtti:1991ae, kulagin2006} yield an effect
that explains only part of the observed EMC effect.  In addition, nuclear
binding models have failed to gain traction in the past, usually due to the
omission of the so-called ``flux factor'' (incorrect treatment of
wave-function normalization)~\cite{li88}, exclusion of pions~\cite{gross92}, and failure
to describe the Drell-Yan data~\cite{Brown:1993sua}. It thus seems unlikely
that the modification of the nucleon pdfs in the nucleus can be explained by
binding effects alone, and aspects of medium modification must be
included~\cite{gross92, benhar97, Miller:2001tg, kulagin2010}.

\begin{figure}[htb]
\centering
\includegraphics[angle=270,width=72mm]{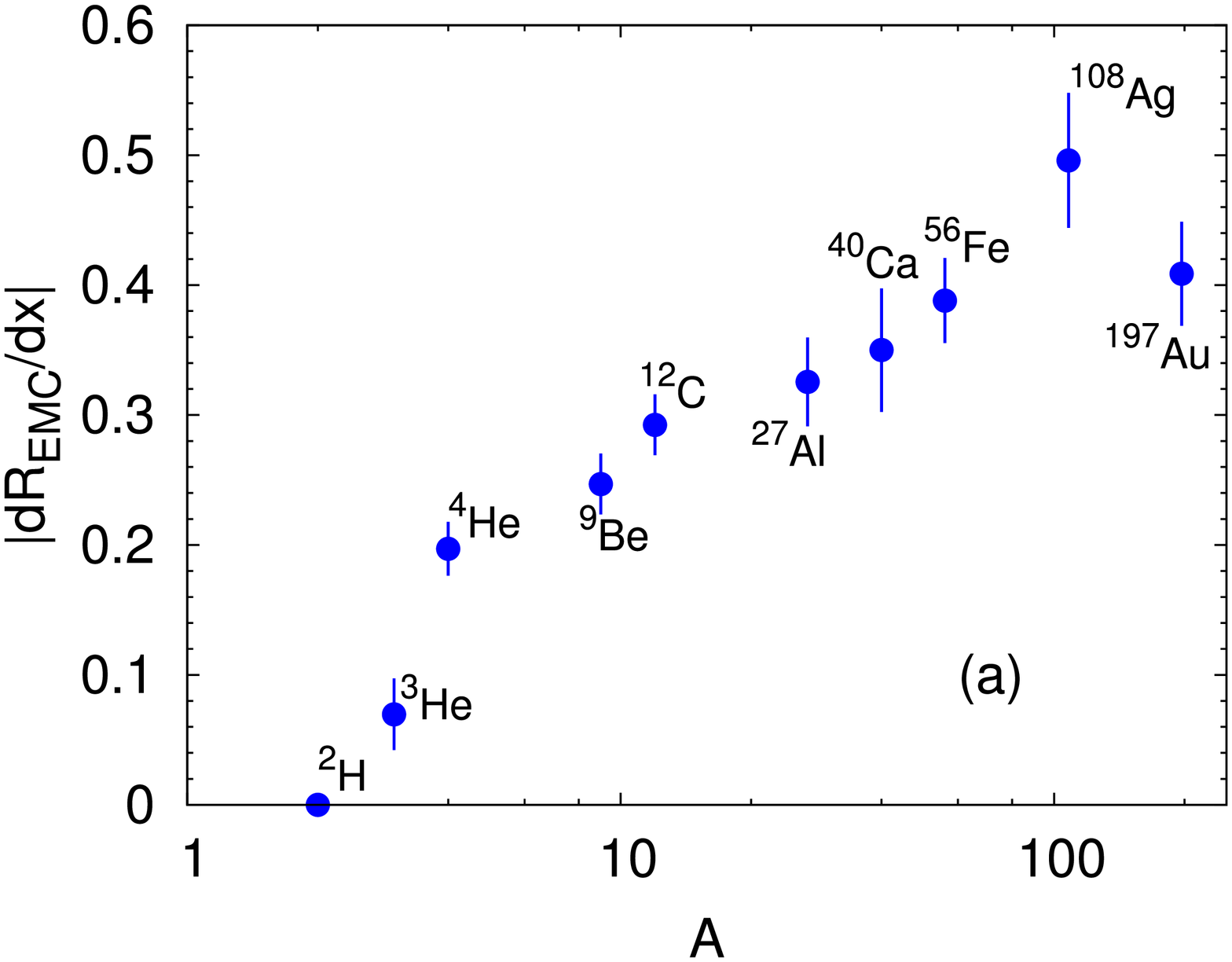}
\includegraphics[angle=270,width=72mm]{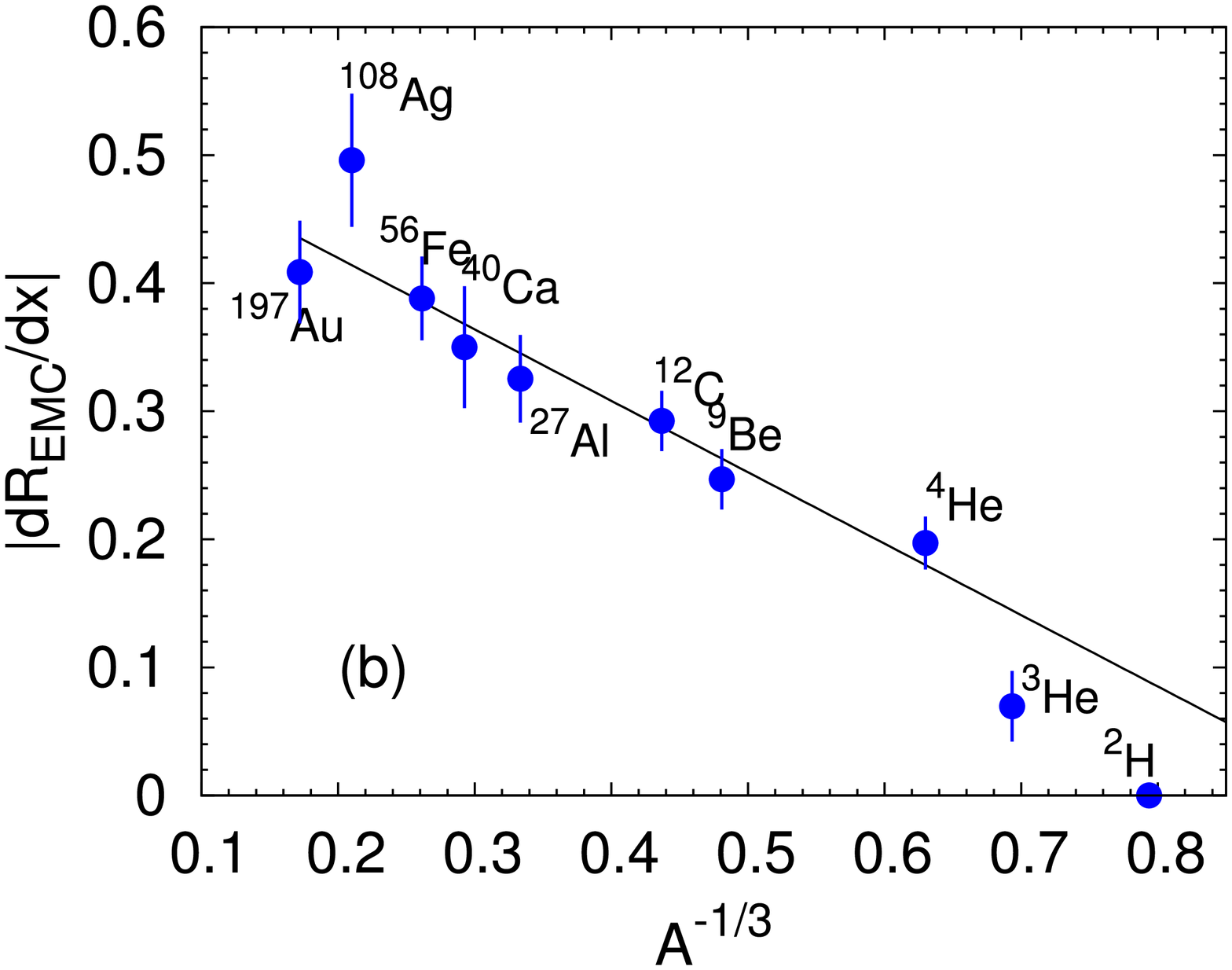}
\caption{(color online) Magnitude of the EMC effect vs. $A$ (top) and
  $A^{-1/3}$ (bottom).
The bottom plot includes a linear fit for $A$$\ge$12.}
\label{fig:emc_nuc_dep_A}
\end{figure}

The E139 analysis~\cite{gomez94} examined the nuclear dependence of the EMC
effect in terms of an ad-hoc logarithmic $A$-dependence and the average
nuclear density. In panel (a) of Fig.~\ref{fig:emc_nuc_dep_A} we show the $A$
dependence.  While it is possible to construct a good linear fit for
either light or heavy nuclei, no linear correlation exists for the
whole data set.

Exact nuclear matter calculations~\cite{Benhar:1989aw} can be applied to finite nuclei
within the local density approximation (LDA)\cite{Antonov86, sick92}.   This provides
an estimate of the $A$ dependence for effects that depend on the nuclear
density and is based on general characteristics of the nuclear density
distributions.  For $A>$ 12, the nuclear density 
distribution $\rho(r)$ has a common shape and has been found to be relatively 
constant in the nuclear interior.  Contributions to the lepton scattering 
cross section from this portion of the nucleus should then scale with $A$.
The nuclear surface is also characterized 
by a nearly universal shape, $\rho(r-R)$, where $R$ is the 
half-density radius $R=r_\circ A^{\frac{1}{3}}$, such that contributions
from the surface grow as $R^2$, or $A^{2/3}$. 
It then follows that the cross section per nucleon (dividing the
separate contributions by $A$) should 
be constant with a small deviation that scales with $A^{-1/3}$, which
is due to the reduced density of the surface region. For small-$A$ nuclei the nuclear response is dominated 
by surface effects while for large-$A$ nuclei the nuclear response 
is dominated by the constant density region.   It has been argued that the response function (per nucleon) for nuclear matter 
can be extrapolated as a linear function of $A^{-1/3}$ to $A^{-1/3} = 0$ 
in the deep inelastic scattering region~\cite{sick92}.

In panel (b) of Figure~\ref{fig:emc_nuc_dep_A} the extracted EMC slope is plotted
versus $A^{-1/3}$.  Somewhat surprisingly, this yields one of the better
correlations with the data, even for $^{12}$C, $^9$Be and $^4$He. This is not
expected, since the prediction of the $A^{-1/3}$ behavior is based on the
assumption of an $A$-independent ``surface'' density distribution and a
scaling with $A$ of the volume/surface ratio.  The assumption that the shape
of the ``surface'' density is universal is certainly not valid for $A\leq$
12, and it is not clear that the concept of dividing the nucleus into a surface region and a
high-density core is at all applicable to $^3$He or $^4$He.

\begin{figure}[htb]
\centering
\includegraphics[angle=270,width=76mm]{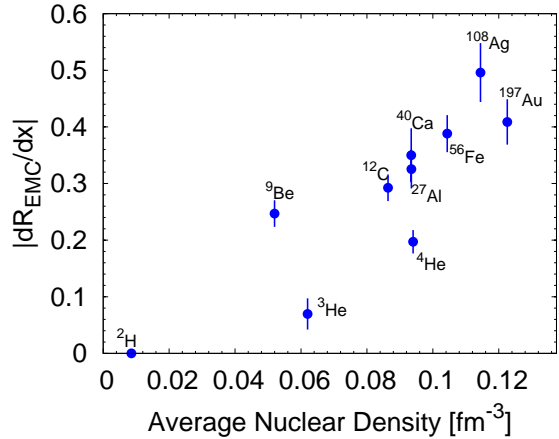}
\caption{(color online) Magnitude of the EMC effect vs. average nuclear density.}
\label{fig:emc_nuc_dep_avgdensity}
\end{figure}

The LDA predicts a simple $A$ dependence based on the assumption that
the EMC effect scales with density.  Since this is not expected to work
for light nuclei, one can evaluate the idea of a density-dependent EMC
effect more directly by looking at the average nuclear
density based on calculations or electron-scattering measurements of the
nuclear mass (or charge) density.
This dependence is shown in Figure~\ref{fig:emc_nuc_dep_avgdensity}.
For light nuclei ($A\le 12$), the average 
density is evaluated using density distributions extracted within Green's
Function Monte Carlo (GFMC) calculations~\cite{Pieper:2001mp,pieper_priv}, while for
heavier nuclei it is derived from electron scattering extractions of the
charge density~\cite{DeJager:1987qc}. This is in contrast to
Ref.~\cite{gomez94}, in which the average density was calculated assuming a
uniform sphere with radius equal to the RMS charge radius of the relevant
nucleus, although for $A\ge12$, this yields the same
qualitative behavior as is seen in
Fig.~\ref{fig:emc_nuc_dep_avgdensity}. 

\begin{figure}[htb]
\centering
\includegraphics[angle=270,width=80mm]{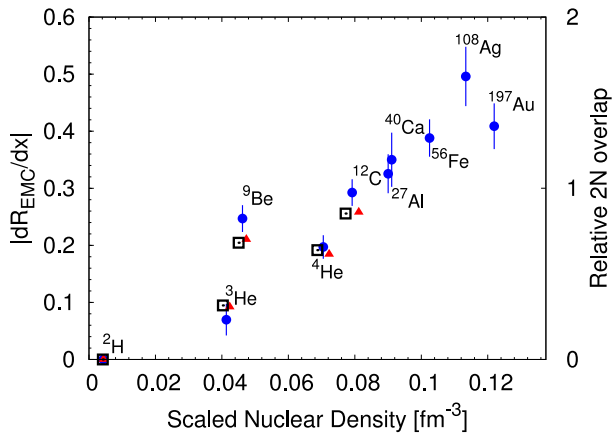}
\caption{(color online) Magnitude of the EMC effect (solid circles) vs. scaled
nuclear density. The solid triangles
and hollow squares show the calculated average 2N overlap from
Eq.~\ref{eq:overlap2} minus the deuteron value, i.e. $\langle O_N \rangle _A - \langle O_N \rangle _D$ (RHS scale: See text for
details). Overlap points are offset on the $x$-axis for clarity.}
\label{fig:emc_nuc_dep_scaleddensity}
\end{figure}

The relationship between EMC slope and density is improved when taking the
scaled nuclear density, which includes an additional correction factor of
$(A-1)/A$, meant to account for the excess nuclear density seen by the struck
nucleon.  This is seen in Fig.~\ref{fig:emc_nuc_dep_scaleddensity}, where the
EMC effect grows approximately linearly with scaled density, with the
exception of $^9$Be.  This was explained in Ref.~\cite{seely09} as being a
result of the cluster--like structure of $^9$Be, whose wave function includes
a sizable component in which the nucleus can be thought of as two $\alpha$
clusters associated with a single neutron~\cite{arai96, Hirai:2010xs,
Pandit:2011zz}. If the EMC effect is governed by the local density, rather
than the average nuclear density, then it is not unreasonable that the size of
the effect in $^9$Be would be similar in magnitude to that in $^4$He.

As mentioned earlier, nucleon wave-functions can have a significant amount of overlap in the
nucleus before the nucleons come close enough to respond to the repulsive core.  If we
can quantify this overlap, it could provide a reasonable measure of the local
density.  We estimate this is by taking the 2-body density distributions from
GFMC calculations~\cite{Pieper:2001mp,pieper_priv} which provide the
distribution of the relative nucleon separation between $pp$, $np$, and $nn$
configurations.  If we integrate the normalized $\rho_2^{pp}(r)$ up to
$r=1.7$~fm, we find the probability that a proton is within 1.7~fm (twice the
RMS radius of a nucleon) of another proton.  Thus, we define a measure of the
relative pair overlap between nucleons by taking
\begin{equation}
O_{NN} = \int_0^\infty{W(r) \rho_2^{NN}(r) d^3r}
\label{eq:overlap1}
\end{equation}
where $W(r)$ is a cutoff function
 used to evaluate the contribution at short distances.  
If $W(r)$ is a step function that cuts off at $r=R_0$, then $O_{pn}$ represents
the average probability that a given $pn$ pair has a separation of $R_0$ or
less.  A proton, then, has an average overlap parameter $O_p = (Z-1)O_{pp} +
NO_{pn}$, which for a step function with $R_0 \to \infty$ yields $(A-1)$, the
total number of neighbor nucleons for the studied proton.  To
obtain the effective 2N overlap for a given reaction, we take a cross section weighted average of
$O_p$ and $O_n$:
\begin{equation}
\langle O_N \rangle = (Z \sigma_p O_p + N \sigma_n O_n) / (Z \sigma_p + N
\sigma_n) .
\label{eq:overlap2}
\end{equation}

We show the relative 2N overlap for two calculations in
Fig.~\ref{fig:emc_nuc_dep_scaleddensity}, subtracting the result for
the deuteron, i.e. $\langle O_N \rangle _A - \langle O_N \rangle _D$.  The solid triangles are for
a step function with $R_0=1.7$~fm and $\sigma_n/\sigma_p=0.5$, although
the result is very insensitive to the exact value of $\sigma_n/\sigma_p$.
Because the amount of overlap between nucleons decreases with the separation,
$W(r)$ can be chosen to enhance the effect when the nucleons are extremely
close together.  The hollow squares are the result when we take $W(r)$ to be a
gaussian centered at $r=0$ with a width of 1~fm.  
 An overall normalization factor is applied to compare
to the $A$ dependence of the EMC slopes. Both of these simple calculations of
overlap yield a good qualitative reproduction of the behavior for light nuclei
and one which is not very sensitive to the choice of the cutoff function or the
exact scale of the cutoff parameter.

For all of
the light nuclei, an average overlap parameter can be obtained from the
\textit{ab initio} GFMC calculations.  This provides realistic input of the
distribution of nucleons in these nuclei, although the quantitative evaluation
of the overlap parameter does depend on the somewhat arbitrary choice of the
cutoff function in Eq.~\ref{eq:overlap1}.  One could use measurements of
short-range correlations in nuclei as an observable which is also sensitive to
the relative contribution from short-distance configurations in nuclei.  This
is a possible interpretation of the correlation observed between SRC
measurements and the EMC effect, and we will present this in detail after
examining the $A$ dependence of the short-range correlation measurements.

To definitively test the notion that the EMC effect depends on ``local
density'', additional data on light nuclei, especially those with significant
cluster structure, are required. Such studies are planned as part of the
program after the Jefferson Lab 12~GeV Upgrade~\cite{E1210008}.

%% file: adep_src.tex
\section{Nuclear Dependence of Short Range Correlations}
\label{adep_src}

Much as DIS isolates scattering from quasi-free quarks, quasielastic (QE)
scattering isolates incoherent scattering from the protons and neutrons in the
nucleus.  This allows us to study the momentum distributions of the bound
nucleons~\cite{benhar08}. Inclusive electron scattering can be used to
pick out
contributions from high-momentum nucleons in SRCs by going to $x>1$
kinematics~\cite{frankfurt93, benhar08, arrington11}.

In the QE regime, we can decompose the cross section into contributions from
single-nucleon scattering (mean-field independent particle contributions) and
scattering from 2-nucleon, 3-nucleon, etc correlations~\cite{frankfurt93}
via:
\begin{equation}
\sigma(x,Q^2) = \sum_{j=1}^{A} A\frac{1}{j}a_j(A) \sigma_j(x,Q^2)\
\label{eq:cssum}
\end{equation}
where $\sigma_j(x,Q^2)=0$ at $x>j$ and the $a_j(A)$'s are proportional to the
probabilities of finding a nucleon in a $j$--nucleon correlation. In the
case of the electron--deuteron cross section, $\sigma_2$ will be
dominated by contributions from 2N correlations for $x>$1.4, where
the nucleon momentum is well above $k_F$ and the mean field
contribution has died off.  In this case, $a_2$ is closely
related to the number of 2N correlations in the nucleus (per nucleon)
relative to that of the deuteron. Hence
Eq.~\ref{eq:cssum} expresses the fact that in the region $j<x<j+1$ the
contribution of $j-$nucleon SRCs dominates. This result is in reasonable
agreement with numerical calculations of the nuclear spectral
functions~\cite{ciofi91, ciofi96}.

\begin{figure}[htb]
 \centering
 \includegraphics[angle=270,width=72mm]{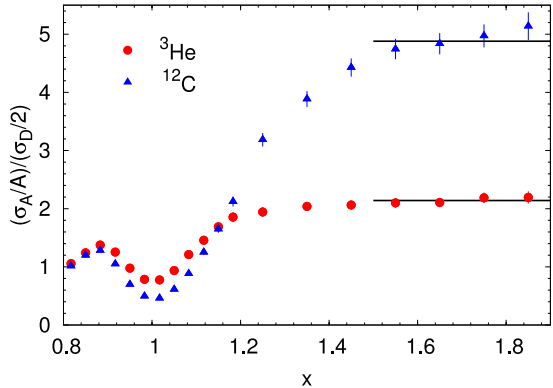} 
 \caption{(color online) Per nucleon cross section ratios for
$^3$He/$^2$H and $^{12}$C/$^2$H measured at JLab~\cite{fomin2012} at
18$^\circ$. In the region dominated by 2N SRCs the ratios becomes
independent of $x$.  The dip around $x$=1 is the result of $A>2$
nuclei having wider quasielastic peaks and the solid line indicates
the region used to extract the ratio $a_2$.}
 \label{fig:src-fomin}
\end{figure}

Equation~(\ref{eq:cssum}) suggests scaling relations between scattering off
the heavy nuclei and the deuteron:
\begin{equation}
\frac{\sigma_A(x,Q^2)/A}{\sigma_D(x,Q^2)/2} = a_2(A)\mid _{ 1.4 \lesssim x \leq 2}
\label{eq:ratios}
\end{equation}
The scaling of the cross section ratios has been established, first at
SLAC~\cite{frankfurt93} and at Jefferson
Lab~\cite{egiyan03,egiyan06,fomin2012}. The most recent experiment measured
this scaling precisely in the 2N correlation region for a range of nuclei with
selected data shown in Fig.~\ref{fig:src-fomin}.

In extracting the relative contributions of 2N SRCs in the inclusive
cross section ratios at $x>1$, it has typically been assumed that the electron
is scattering from a pair of nucleons with large relative momentum
but zero total momentum, such that the cross section for scattering from
a neutron-proton pair in a nucleus is identical to the cross section for
scattering from a deuteron.  In this case, the elementary
electron--nucleon cross sections as well as any off-shell effects cancel in
taking the ratio.  Final state interactions are also assumed to cancel in
the cross section ratios~\cite{frankfurt93, arrington11}.

Earlier analyses~\cite{frankfurt93,egiyan03,egiyan06} assumed that the SRCs
would be isospin-independent, with equal probability for $pp$, $np$, and $nn$
pairs to have hard interactions and generate high-momentum nucleons. This
necessitated an ``isoscalar correction'' to account for the excess of $nn$ (or
$pp$) pairs in non-isoscalar nuclei as well as the difference between the
$e-p$ and $e-n$ elastic cross sections.  More recently, measurements of
two-nucleon knockout showed that these correlations are dominated by $np$
pairs~\cite{piasetzky06, subedi08} due to the fact that the bulk of the
high-momentum nucleons are generated via the tensor part of the N--N
interaction rather than the short-range repulsive core~\cite{sargsian05,
schiavilla06}. The most recent experiment~\cite{fomin2012} to precisely
measure SRCs on a range of nuclei did not apply this isoscalar correction, and
presented results for previous measurements with this correction removed.

\begin{figure}[htb] 
 \centerline{\includegraphics[angle=270, width=72mm]{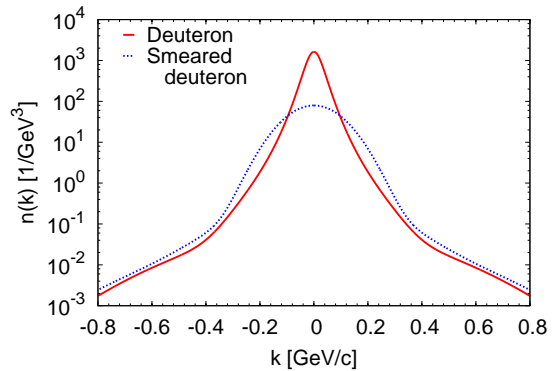}}
 \caption{(color online) Momentum distribution for the free deuteron and an
 $np$ pair in iron, taken as the sum of a mean field
 (gaussian) contribution and the convolution of the high-momentum
 deuteron tail with the
 c.m.~motion of the pair in iron.}
 \label{fig:cmmotion}
\end{figure}

The per nucleon cross section ratio at large $x$ provides a direct measure of
the contribution of high-momentum nucleons relative to the deuteron.  However,
this is not equal to the relative number of SRCs, since in $A>2$ nuclei, the
correlated pair experiences motion in the mean field created by the rest of
the nucleons.  The momentum distribution of the pair will be smeared out, which
will flatten the top of the QE peak, depleting the low-momentum part of the
distribution, but enhancing the high-momentum tail.  The effect is illustrated
in Fig.~\ref{fig:cmmotion} which shows the deuteron momentum distribution
along with an estimate of the momentum distribution for an $np$ pair in iron.
The ``smeared deuteron'' curve is generated by taking the high-momentum part of
the deuteron distribution and convolving it with a pair
c.m.~distribution to estimate the impact of the motion of the correlated pair
in the nucleus.  This is combined with a gaussian distribution whose width is
chosen to reproduce a mean field calculation for iron~\cite{ciofi96}, and whose
magnitude is such that the total distribution is properly normalized.

A correction for this redistribution of strength was first applied in
Ref.~\cite{fomin2012}, where analyses of previous experiments were
also updated.  The correction procedure was based on the calculation
of Ref.~\cite{ciofi96}, where the deuteron momentum distribution was convolved
with a parametrization of c.m.~motion of the pair, which yielded a 20\%
enhancement in the high momentum tail for iron. This correction was applied to the other nuclei by assuming
that the enhancement in the ratio, which scales with the c.m.~momentum of the
pair, was proportional to the Fermi motion of the nucleus.

We performed a similar convolution calculation for a variety of nuclei, and
found a slightly larger correction, above 30\% for iron.   We also
observed that the size of the effect depends on the momentum region examined,
the details of the deuteron momentum distribution and the assumed c.m.~momentum
distribution.  While the enhancement is relatively constant at large $k$, it does
increase for very large momenta. This effect is at least partially
responsible for the small rise of the SRC ratios for $x \to 2$.

Another recent attempt to estimate the role of c.m.~motion~\cite{Vanhalst:2012ur}
yielded significantly larger corrections, although it is not yet clear
what explains this difference. In all of the above cases, the correction has only a
very weak nuclear dependence, mostly yielding an overall scaling
factor for the SRC ratios, without significant impact on the linear
correlation between them and the EMC effect.  We use the correction and uncertainty applied in Ref.~\cite{fomin2012}, but it is clear that this is an issue
requiring further study.

\begin{table}[htb]
\begin{center}
\caption{ Existing measurements of SRC ratios, $R_{2N}$, all corrected for
c.m.~motion of the pair and excluding the isoscalar correction applied to
earlier extractions.  The second-to-last column combines all the measurements,
and the last column shows the ratio $a_2$, obtained without applying
the c.m. motion correction.  SLAC and CLAS results do not have
Coulomb corrections applied, which would raise the CLAS Fe ratio by
$\sim$5\%, and the SLAC Au data by $\sim$10\% (correction
is kinematic-dependent).}
\label{src_table}
\begin{tabular}{|c|c|c|c|c|c|}
\hline
        & E02-019 & SLAC & CLAS & $R_{2N}$-ALL&a$_2$-ALL\\ \hline
$^3$He  & 1.93$\pm$0.10 & 1.8$\pm$0.3   & --            & 1.92$\pm$0.09&2.13$\pm$0.04\\
$^4$He  & 3.02$\pm$0.17 & 2.8$\pm$0.4   & 2.80$\pm$0.28 & 2.94$\pm$0.14&3.57$\pm$0.09\\
Be      & 3.37$\pm$0.17 & --            & --            & 3.37$\pm$0.17&3.91$\pm$0.12\\
C       & 4.00$\pm$0.24 & 4.2$\pm$0.5   & 3.50$\pm$0.35 & 3.89$\pm$0.18&4.65$\pm$0.14\\
Al      & --		& 4.4$\pm$0.6	& -- 		& 4.40$\pm$0.60&5.30$\pm$0.60\\
Fe      & -- 		& 4.3$\pm$0.8	& 3.90$\pm$0.37 & 3.97$\pm$0.34&4.75$\pm$0.29 \\
Cu      & 4.33$\pm$0.28 & -- 		& -- 		& 4.33$\pm$0.28&5.21$\pm$0.20 \\
Au      & 4.26$\pm$0.29 & 4.0$\pm$0.6   & --            & 4.21$\pm$0.26& 5.13$\pm$0.21 \\
\hline
\end{tabular}
\end{center}
\end{table}

For the purposes of our analysis, we combine the results of the JLab
Hall-C~\cite{fomin2012}, Jlab Hall-B (CLAS)~\cite{egiyan06} and
SLAC~\cite{frankfurt93} measurements. The combined data set provides a large
collection of nuclei to examine the $A$ dependence of the extracted SRC
contributions.  Table~\ref{src_table} shows $a_2$, the raw $A/D$ cross
section ratio, as well as $R_{2N}$, where the c.m.~motion correction
has been applied.   The meanings of the two quantities are subtly different. 
$a_2$ represents the relative strength of the high-momentum
tail, i.e. the total contribution from high momentum nucleons relative to the
deuteron.  For iron, $R_{2N} \approx 4$, implying that a nucleon in iron is
four times more likely to be part of an SRC than a nucleon in a deuteron,
At the same time $a_2$ is  $\approx 4.8$, which means that there are
almost five times as many high momentum nucleons in the tail of the
distribution for iron as there are for the deuteron.  This 20\%
enhancement comes about due to the c.m.~motion of the correlated
pair. 

\begin{figure}[htb] 
 \centerline{\includegraphics[angle=270, width=76mm]{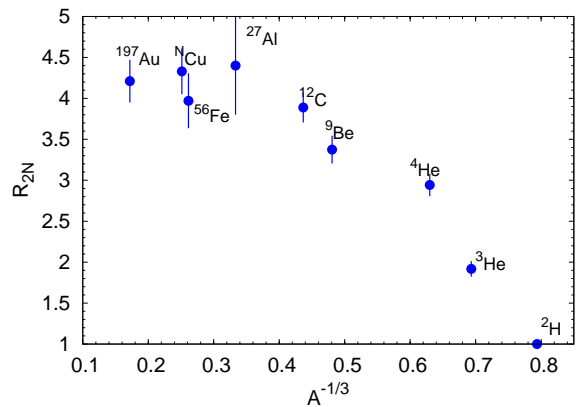}}
 \caption{(color online) $R_{2N}$ versus $A^{-1/3}$. }
 \label{fig:AdepSRC}
\end{figure}

We use $R_{2N}$ for the nuclear dependence tests, as we are examining the
behavior of the number of SRC pairs relative to the deuteron. Since
the c.m.~correction factor applied to $a_2$ has very little $A$
dependence, the nuclear dependence $a_2$ is very similar to that of $R_{2N}$.

Figure~\ref{fig:AdepSRC} shows $R_{2N}$ as a function of
$A^{-1/3}$, the behavior expected in the LDA~\cite{Antonov86,sick92}.  While
$R_{2N}$ is a relatively smooth function of  $A^{-1/3}$, there is
not a simple, linear relation suggesting a proportionality.  As with
the EMC effect, the prediction of scaling with $A^{-1/3}$ is an
approximation which is not expected to be valid for very light nuclei.

\begin{figure}[htb] 
\centerline{\includegraphics[angle=270,width=83mm]{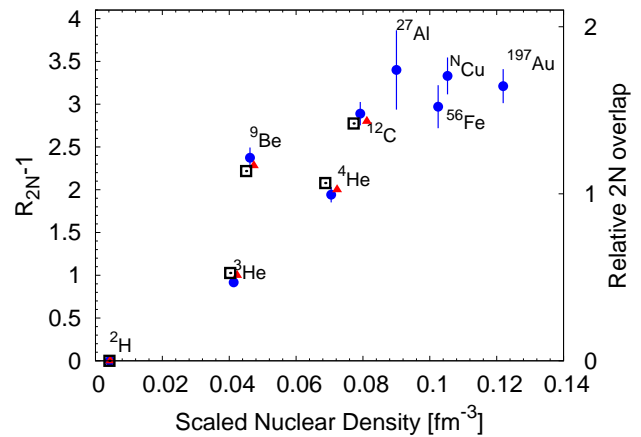}}
\caption{(color online) $R_{2N}$ versus scaled nuclear density
(solid circles). The solid triangles and hollow squares show the calculated 2N
overlap minus the deuteron value (RHS scale) from Eq.~\ref{eq:overlap2}.
Points offset on the $x$-axis for clarity.}
\label{fig:rhodepSRC}
\end{figure}

For nuclei with similar forms for $\rho(r)$, we expect to see scaling of the
SRCs, that is, denser nuclei are more likely to have short range
configurations. Figure~\ref{fig:rhodepSRC} shows $R_{2N}$ as a function of
the scaled nuclear density, defined in the previous section. It is clear that
the simple density-dependent model does not track the behavior of the light
nuclei, whose large deviations are reminiscent of those shown by the EMC
effect~\cite{seely09}.

While there were few calculations for the nuclear dependence of the EMC effect that went beyond a simple scaling with $A$ or density, there are more calculations for
the $A$ dependence of short-range correlations. The authors of Ref.~\cite{mcgauley11} estimate the relative SRC contribution in a variety of nuclei based on the mean field densities
of the nuclei. Ref.~\cite{Vanhalst:2011es} estimates the high-momentum contributions based on the presence of deuteron-like pairs in the nucleus.  In both of these
approaches, they predict a stronger rise of the SRC contributions with
$A$ than is observed in the data.  The latter result~\cite{Vanhalst:2011es}
displays a sensitivity to the c.m.~motion correction~\cite{Vanhalst:2012ur}, yielding
an even sharper rise with $A$.  An earlier work~\cite{sato86} estimated the
probability for multi-quark ($6q$, $9q$,...) clusters in nuclei based on the probabilty of overlap of two or more nucleons within some critical separation, which should
also be closely related to the contribution of 2N and 3N-SRCs.  They
find that the probability for $6q$ configurations scales roughly with
the density of the nucleus, with the exception of $^4$He, where a much
larger contribution is predicted.  In the data, the SRC ratio in $^9$Be is significantly
larger than expected for a model which scales with density, and $^4$He
does not show the anomalously large contribution predicted by this
calculation.

As with the analysis of the EMC effect, we also show the
relative 2N overlap from Eq.~\ref{eq:overlap2} as a function of the
scaled nuclear density in Figure~\ref{fig:rhodepSRC}.
 As before, the overlap for the deuteron has
been subtracted for $A>$2.  The density dependence of SRCs is well
reproduced by the overlap calculation, as they both reflect the
abundance of short-distance configurations.  This allows us to use SRC
ratios as an experimental measure of overlap, extending the comparison
with the EMC effect in Fig.~\ref{fig:emc_nuc_dep_scaleddensity} to
$A>$12.  Such a comparison assumes a certain connection between SRCs
and the EMC effect, which will be examined in detail in the next
section, along with an alternate possibility.

%% file: src_vs_emc.tex
\section{Detailed Comparison of SRC and EMC results}

\begin{figure}[htb]
\includegraphics[angle=270,width=78mm]{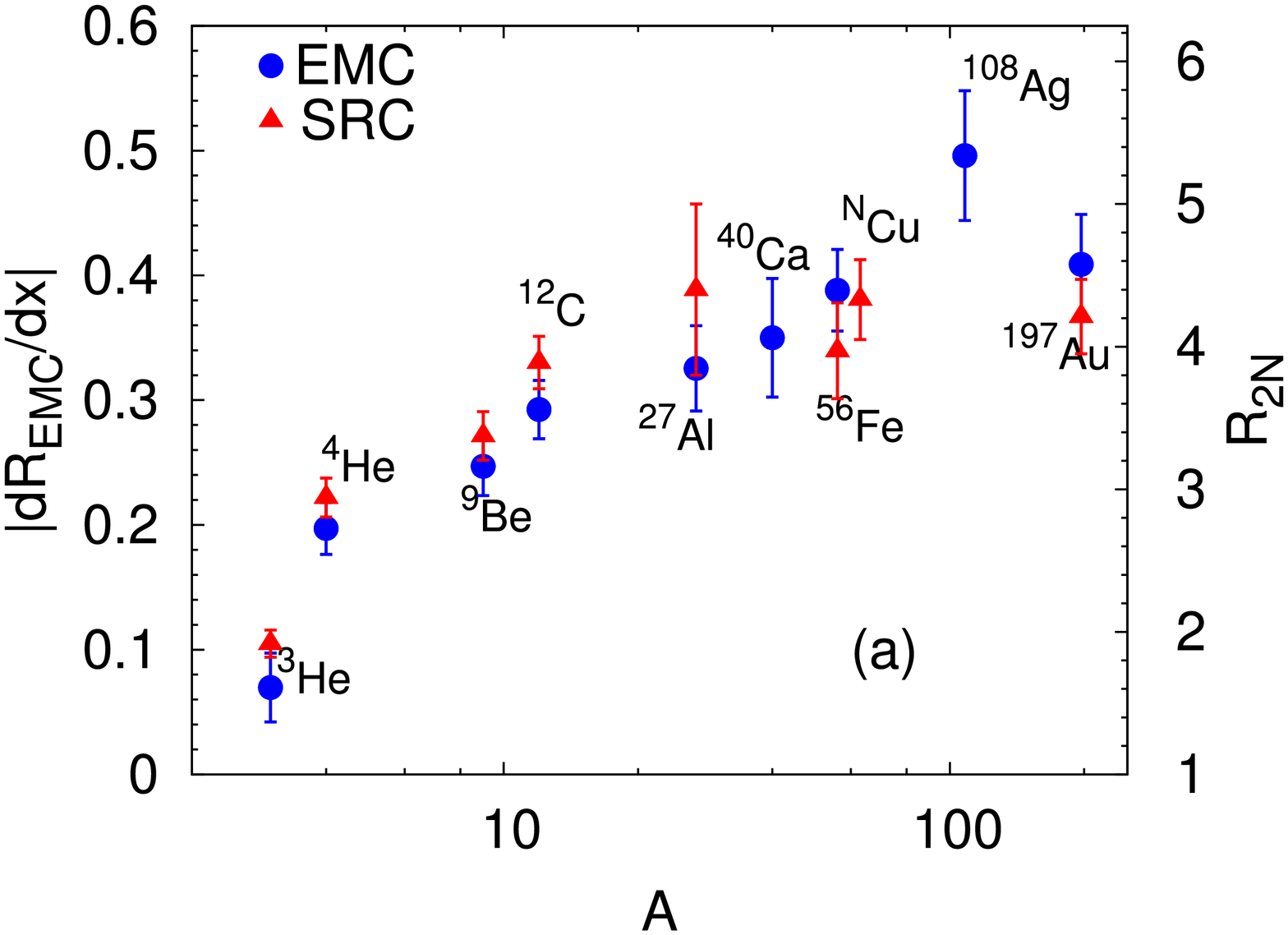} \\
\includegraphics[angle=270,width=78mm]{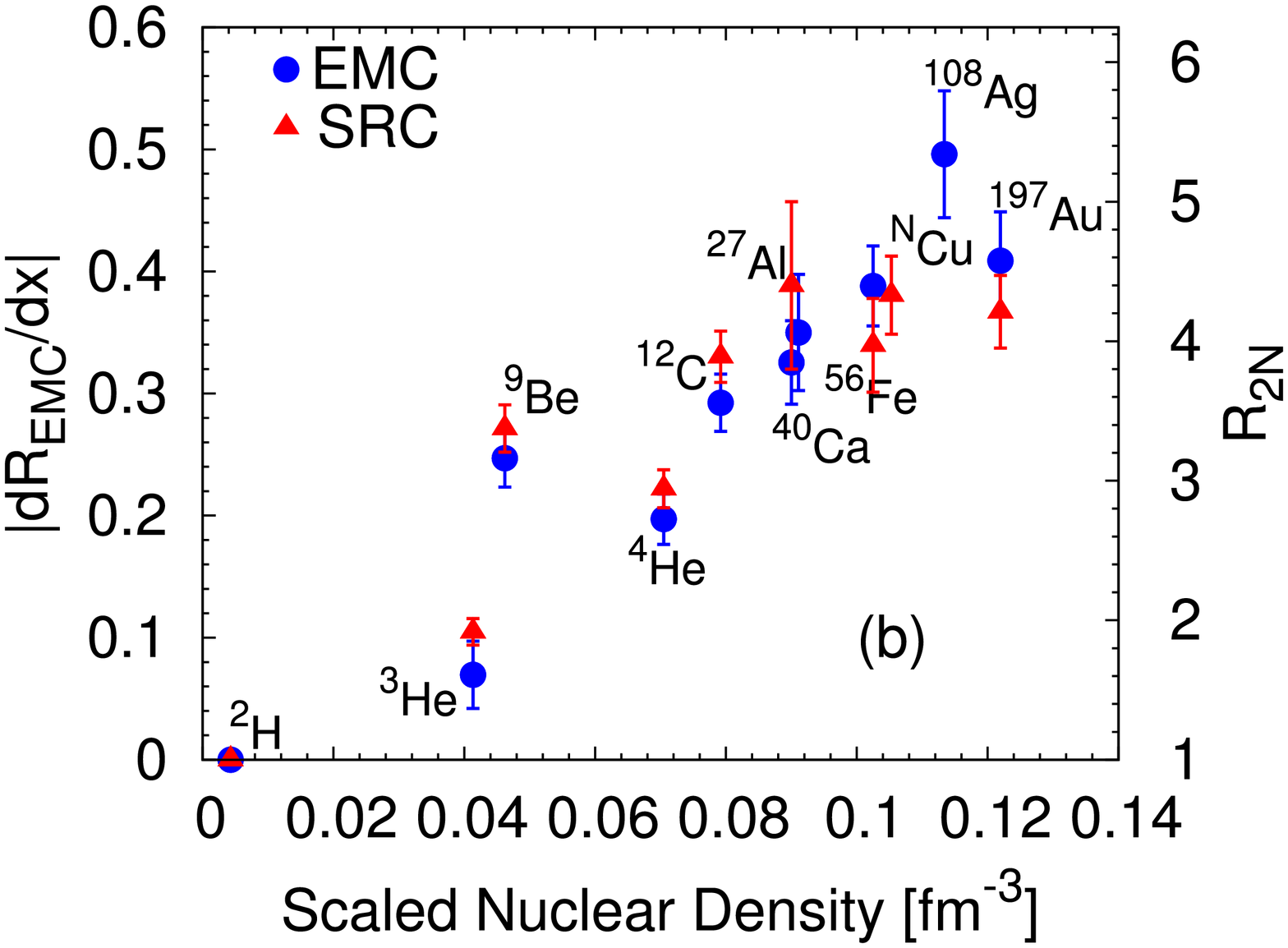}
\caption{(color online) Size of the EMC effect ($|dR_{EMC}/dx|$ as
  well as the relative measure of SRCs ($R_{2N}$) are shown as a
  function of $A$ (top) and scaled nuclear density (bottom).}
\label{fig:src_emc_light}
\end{figure}

As discussed in the introduction, there have been previous comparisons of
the nuclear dependence of the size of the EMC effect and the contributions from
SRCs in nuclei~\cite{weinstein2010rt, Hen:2012fm}.  Given the data available in
the initial analysis, the correlation seen between the two effects could be
explained by a common density- or $A$-dependent scaling.  However, the new
data on the EMC effect~\cite{seely09} and SRCs~\cite{fomin2012} rule out this
simple explanation, while exhibiting almost identical trends versus both
density and $A$, shown in Fig.~\ref{fig:src_emc_light}.  For
the EMC effect, it was suggested that if the local environment of the
struck nucleon drives the modification of the quark distributions, then the
strong contribution of $\alpha$-like clusters would make $^9$Be behave like a
much denser nucleus.  The nearly identical behavior of $^9$Be in the SRC
extraction~\cite{fomin2012} supports this idea, as the SRC measurements
directly probe the short-distance structure.  Even with these new data and
their unexpected but common trend, the linear relationship observed
in~\cite{weinstein2010rt} is still apparent~\cite{Hen:2012fm}. This suggests
that a careful re-examination of the linear correlation is in order to try to
better understand its underlying cause.

First, we note that the initial comparison of the EMC effect and 
SRCs used extractions of the SRCs which included an isoscalar
correction for nuclei with unequal numbers of protons and neutrons and did not
apply corrections for c.m.~motion of the correlated pair. It has been
shown that SRCs are made-up of predominantly $np$ pairs due to the
tensor interaction~\cite{sargsian05, schiavilla06,
subedi08}, making the isoscalar correction unnecessary.  
The question of the c.m.~motion correction is
somewhat more complicated in the context of the
direct comparison of the EMC and SRC results.  Whether or not this correction
should be applied in this analysis depends on exactly what correlation is
being examined, and so we focus now on the different explanations for this
correlation.

The fact that $^9$Be so obviously violates the density dependence for both
effects in the same way  suggests that an altered density dependence, such as
``local density'' (LD) may give us a good description of both effects. One
should then compare the size of the EMC effect to $R_{2N}$, which represents
the relative probability that a nucleon will be part of a very short-distance
configuration (a deuteron-like SRC).  For extremely small nucleon
separations, the short-range repulsive core will yield hard
interactions, and thus high momenta, for all $NN$ pairs.  However, the bulk of the SRCs observed
are $np$ pairs~\cite{subedi08}, generated by the longer-range tensor interaction. If $nn$, $np$, and $pp$ pairs all have
equal probability to form high local density configurations, we would expect
that the EMC effect should scale with the number of possible NN pairs in the
nucleus, $N_{tot}=A (A-1)/2$, while the SRC contribution is sensitive to only
the possible $np$ pairs, $N_{iso}=N Z$.   For light nuclei, we test this
assumption by examining the two-body density distributions~\cite{Pieper:2001mp,pieper_priv} for all $NN$ pairs.
Taking nuclei up to $A$=12, the $np$ pairs have a larger probability to have
small separation, on average 10-20\% more than for $pp$ or $nn$ pairs at
separations below 1.7~fm.  So while the assumption that all pairs contribute
equally at short distances isn't exact, it's significantly better than
assuming that only $np$ pairs contribute.  Thus, we scale the SRC ratio by a
factor $N_{tot}/N_{iso}$ to account for the difference in the pair counting
for the EMC and SRC data, which is a simple first-order correction
that neglects the possible impact of nuclear structure effects.

\begin{figure}[htb] 
\includegraphics[angle=270, width=72mm]{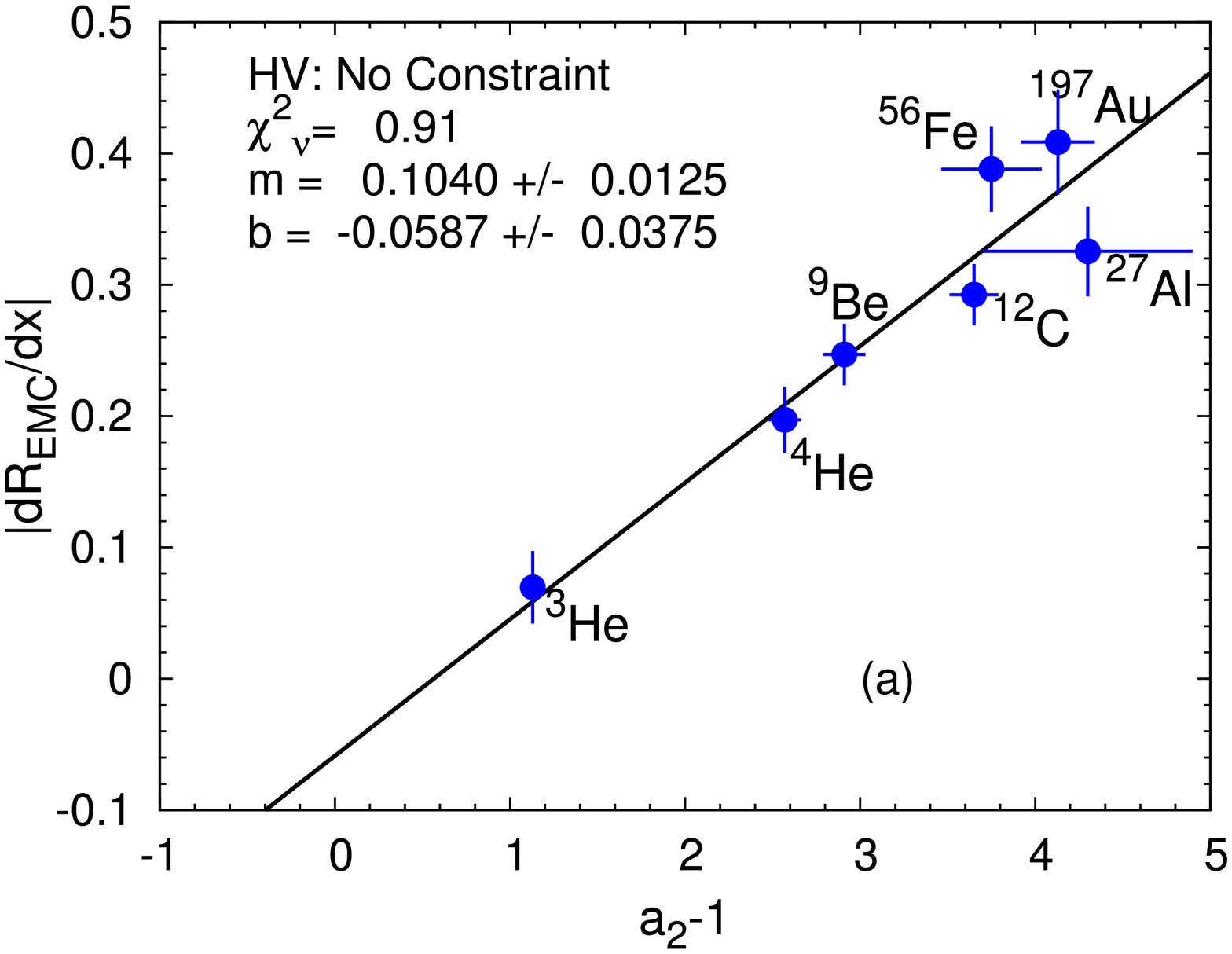}
\includegraphics[angle=270, width=72mm]{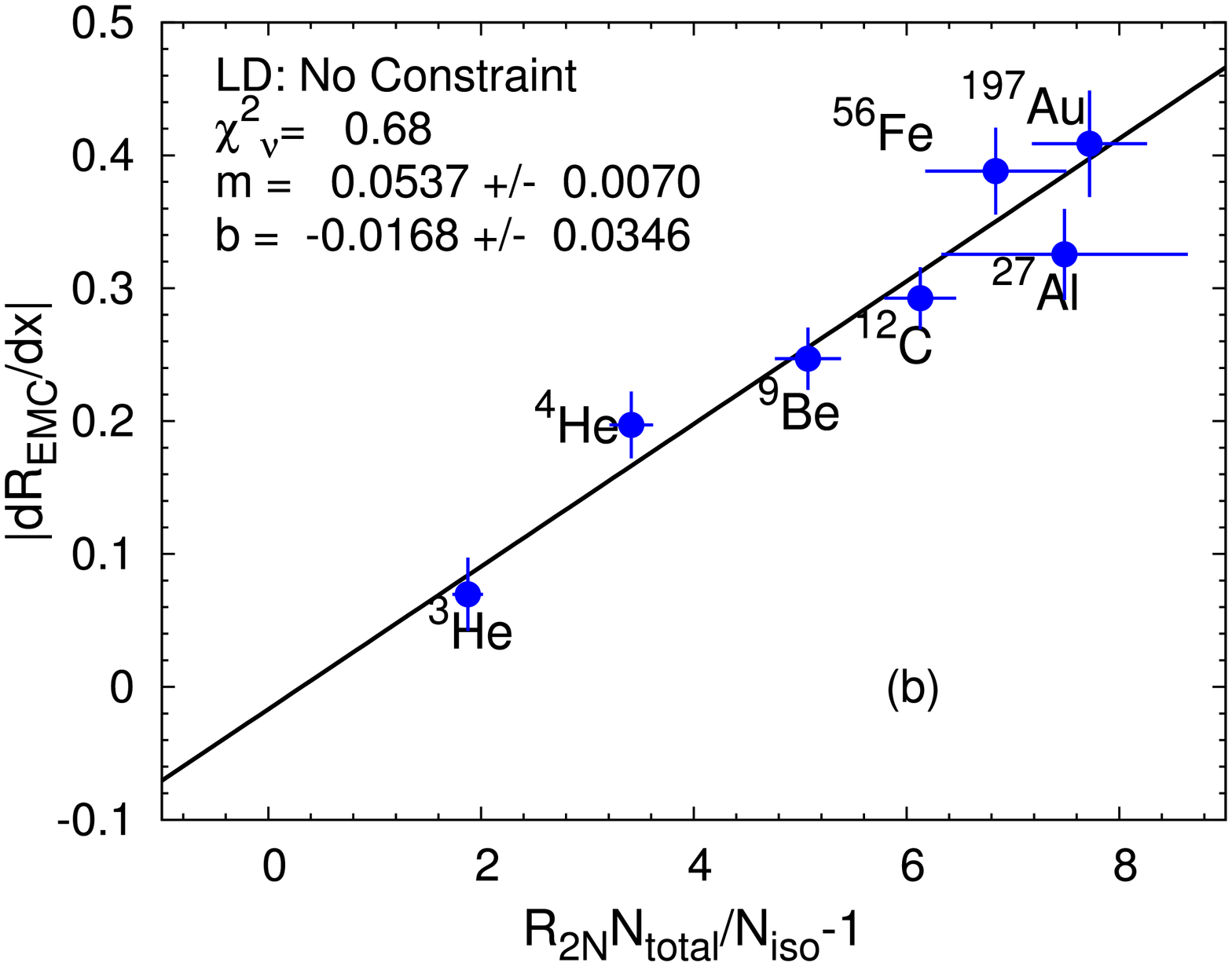}
\caption{(color online) Comparison of EMC slopes and SRC observables from world's
data where both observables are available for the same nuclei.  The top plot shows
the EMC slope vs. $a_2$, testing the high virtuality interpretation. 
Analogously, the bottom plot shows the EMC slope vs. $R_{2N}$ scaled
by $N_{tot}/N_{iso}$ and normalized to the deuteron, testing the local density
interpretation (as discussed in the text) along with fits.}
\label{src_vs_emc_raw}
\end{figure}

A different hypothesis to explain the linear relationship between the two
effects was proposed by Weinstein et al~\cite{weinstein2010rt},
suggesting that the EMC effect is driven by the virtuality of the
high-momentum nucleon~\cite{sargsian03, ciofi2007}.  In this case, it is
the relative probability for a nucleon to have high momentum ($>k_{F}$) that should
drive the EMC effect, and thus the uncorrected $a_2$ SRC ratio is a more direct
indicator of the underlying explanation.  We will refer to this
hypothesis as ``high-virtuality'' (HV).

We now make two comparisons to examine the relationship between
the EMC effect and SRCs using these two different underlying assumptions.
The data as well as the linear fits for both approaches are shown in
Fig.~\ref{src_vs_emc_raw}. A
two-parameter linear fit is performed for both cases without
any constraint for the deuteron. Thus, we can examine the fit to test
both the linear correlation of
the observables and the extrapolation to the expected deuteron value. The
intercept of the fit is expected to be zero, since both the EMC effect and SRC
contributions are taken relative to the deuteron.

Both approaches yield reasonable results, but
we have to delve into the details to understand the impact of the
small differences.  While the LD fit has a better $\chi^2_\nu$ value,
the fractional errors of the points of the $x$-axis are larger due to
the additional model-dependent uncertainties arising from the c.m.~motion
correction~\cite{fomin2012}. A 30\% uncertainty was applied for this
$A$-dependent correction, but any error in this correction is likely to have
a smooth $A$ dependence, so treating these as uncorrelated will
artificially lower the $\chi^2_\nu$ value.  If we repeat the LD fit in
panel (b) of Fig.~\ref{src_vs_emc_raw} neglecting this extra model-dependent uncertainty
(i.e. taking the same fractional uncertainty on $R_{2N}$ as we use for $a_2$),
the reduced $\chi^2_\nu$ value increases from 0.68 to 0.83, as compared to
0.91 for the HV fit.  Overall, the LD fit appears to do a better job: the
extrapolation of the fit to the deuteron gives essentially zero, as it
should, and it has a smaller $\chi^2$ value.  However, neither of these
differences significantly favors the LD hypothesis.

\begin{figure}[h!] 
\includegraphics[angle=270, width=72mm]{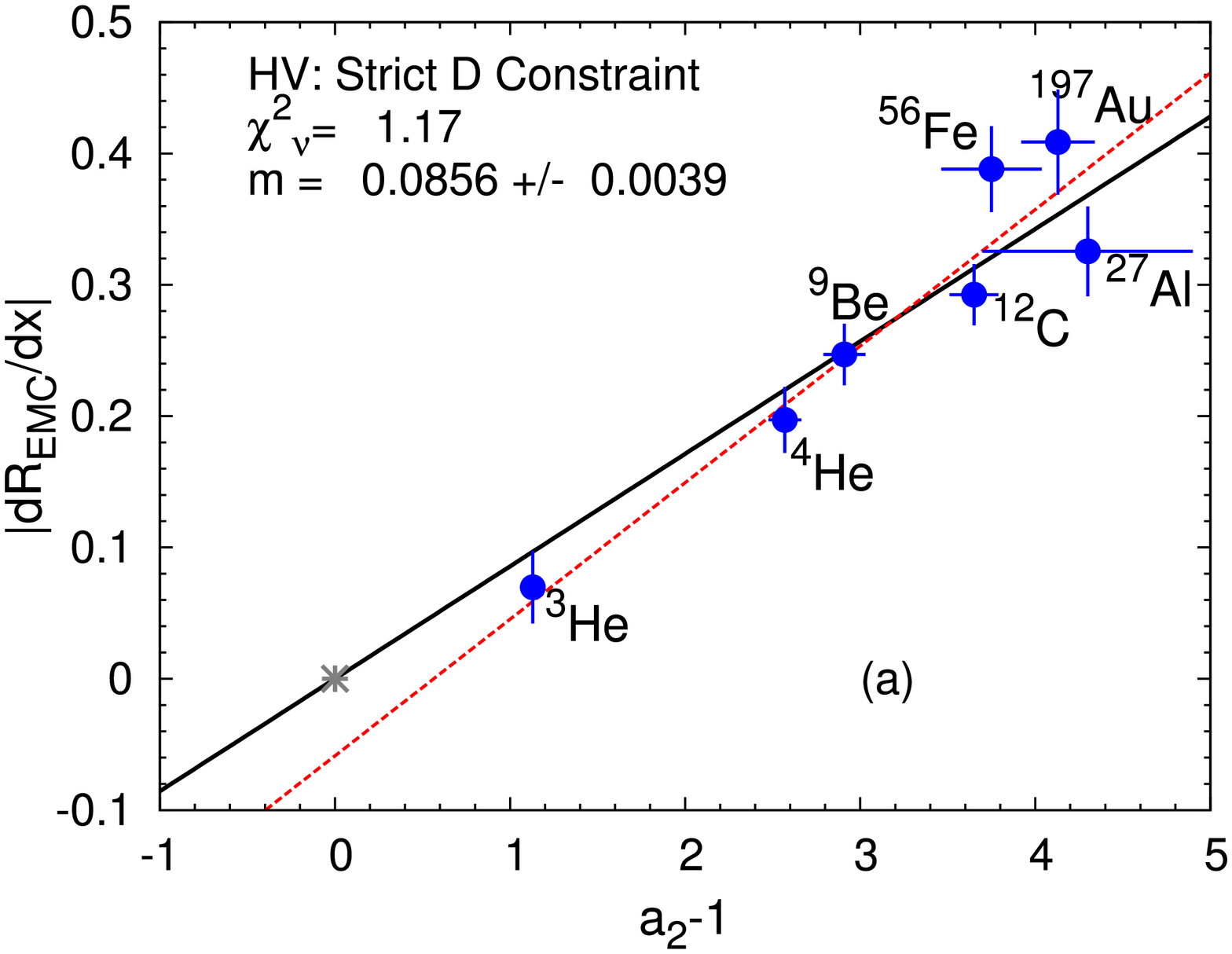}
\includegraphics[angle=270, width=72mm]{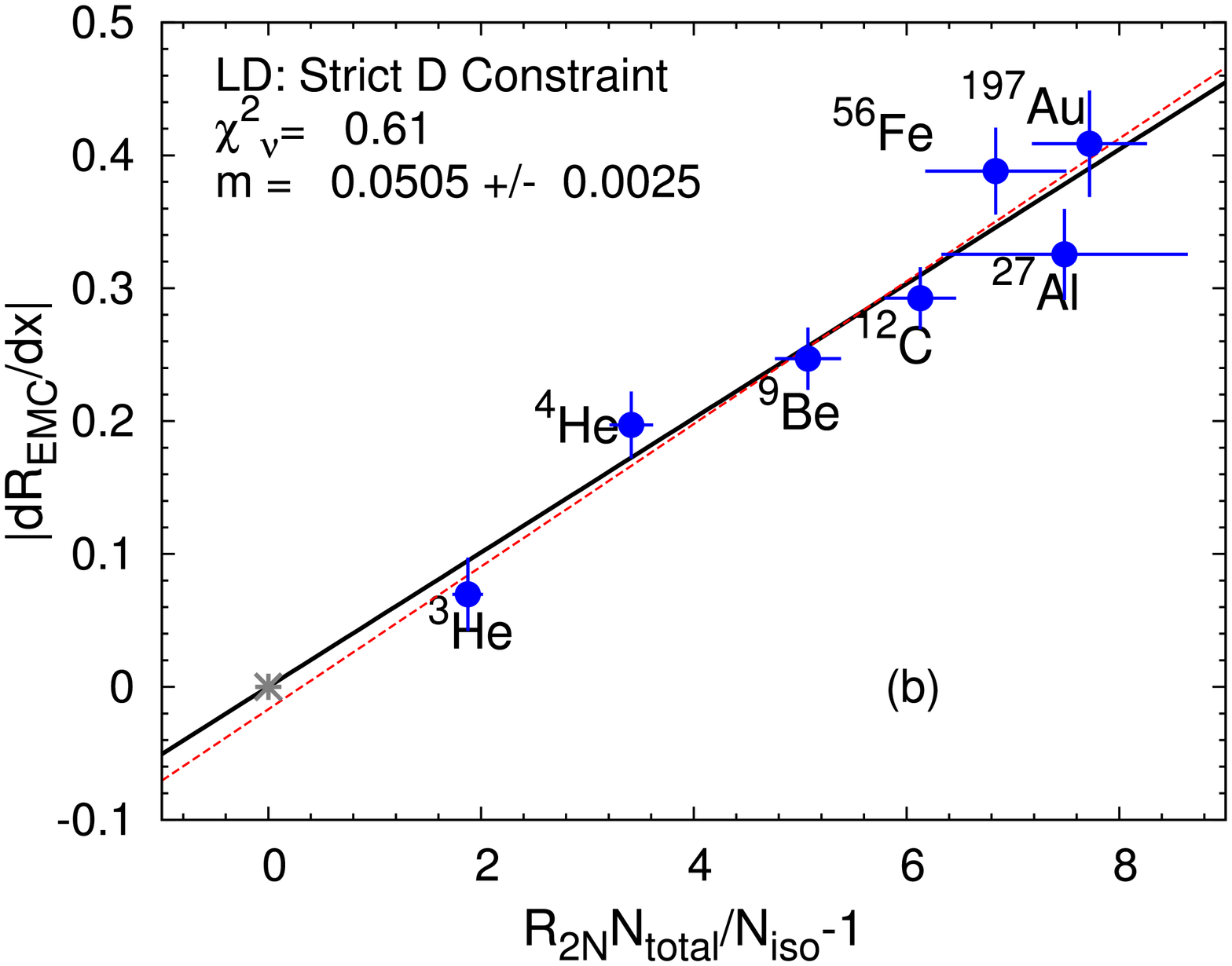}
\caption{(color online) EMC slopes vs $a_2$ (top) and  $R_{2N}$ scaled
by $N_{tot}/N_{iso}$ and normalized to the deuteron (bottom).
The solid line is the one-parameter fit, constrained to yield zero for
the deuteron (grey point).  The dashed red line shows the result of the
two-parameter (Fig.~\ref{src_vs_emc_raw}) for comparison. The fits
are almost indistinguishable for the LD tests.}
\label{src_vs_emc_raw_force0}
\end{figure}

Next, we remove the intercept as a free parameter (leaving only the
slope), and thus constrain the fit by forcing the EMC effect to go through
zero for the deuteron ($a_2$=$R_{2N}$=1).  The $\chi^2_\nu$ of this fit should
test both the linearity and the consistency with the deuteron, allowing for a
more quantitative comparison of the results.  For the constrained fit, these
can be seen in Fig.~\ref{src_vs_emc_raw_force0}. The gap in the $\chi^2_\nu$
values for the two approaches grows, with 1.17 for HV and 0.61 (0.73 when
taking fractional uncertainties from HV case) for LD fits, corresponding
to a total change $\Delta \chi^2$=3.4 (2.6). While the LD
interpretation yields a better description of the data, $\chi^2_\nu=1.17$ for
the HV fit corresponds to a 32\% confidence level, so the data are consistent
with either hypothesis.

The HV approach with the single parameter fit (HV-0) most closely reflects the
previous analysis~\cite{weinstein2010rt, Hen:2012fm}, in that they used $a_2$
as the measure of SRCs and the raw EMC effect slope. However, the quantitative
results of the two analyses differ due to the inclusion of different sets and
corrections factors applied to the data.  The initial work used the older
extractions of $a_2$~\cite{egiyan06} which applied the isoscalar correction
that we now know is not appropriate. The correction was applied to all of the
experimentally measured $A$/$^3$He ratios, which were combined with a
$^3$He/$^2$H ratio based largely on a calculation which did not include the
isospin correction for $^3$He.  Therefore, even the isoscalar $A$/$^2$H ratios
end up with this correction applied.  The authors of
Ref.~\cite{Hen:2012fm}  make several
independent extractions of the correlation, comparing the
EMC results to the original CLAS SRC ratios,
as well as the updated JLab Hall C results~\cite{fomin2012}.  They also
compare different versions of the Hall C data, using both $a_2$ and $R_{2N}$,
and also examining $a_2$ without the coulomb corrections but with the
old-style isoscalar corrections, which more closely reflects the
analysis of the CLAS data.  Note that in the comparison to the Hall C
data, they compare the copper SRC ratios to the EMC effect measurements for
iron.  We use the data shown in Tables~\ref{tab:2n_test} and~\ref{src_table} for common nuclei only.

While the one-parameter fit is a useful way to compare the relative quality of
fits for the LD and HV inspired foundations, it yields an unrealistic estimate
for the uncertainties on the fit.  Including a
deuteron constraint point neglects the fact that there are significant
correlated uncertainties in all of the EMC or SRC points from a single
experiment, since all of the values are measured relative to the deuteron.
Therefore, the statistical and systematic uncertainties in the deuteron data
generate an overall normalization of the values for all other nuclei from that
measurement which is neglected entirely in this approach. In addition,
the linear fit will have extremely small uncertainties for nuclei close to
the deuteron, yielding fit uncertainties for light nuclei (or the extrapolation
to the free nucleon) that are significantly smaller than for any existing
measurement.

We can evaluate the impact of this and make a more realistic estimate of the
fit uncertainties by adding a deuteron constraint point which includes a
reasonable estimate of the uncertainty associated with the deuteron
measurements in the experiments.  We take $|dR_{EMC}/dx|=0\pm0.01$, and
$a_2 = R_{2N}=1 \pm 0.015$, where the error bars were estimated based on
deuterium cross section uncertainties from Refs.~\cite{seely09}
and~\cite{fomin2012}. The extracted slope is almost unchanged, while the
uncertainty increases by approximately 20\%.

\begin{table}[htb]
\begin{center}
\caption{Summary of linear fits of EMC effect vs $R_{2N}$ or $a_2$, and
extrapolations to the slopes of the EMC effect for the deuteron, EMC(D), and IMC effect
for the deuteron, IMC(D).''-0'' denotes a
1-parameter fit, forcing the line to go through zero, corresponding to no EMC
effect for the deuteron. ``-D'' denotes a two parameter fit  including
a realistic deuteron constraint described in the text.  The number in parentheses of the $\chi^2_\nu$ column includes the
result of fitting with smaller fractional errors from $a_{2}$.}
\vspace*{0.25in}
\begin{tabular}{|c|c|c|c|}
\hline
As Published & $\chi^2_\nu$ & EMC(D) & IMC(D)\\
\hline
\hline

HV (Fig.~\ref{src_vs_emc_raw})          &~0.91       & -0.0587$\pm$0.037 & 0.1040$\pm$0.012 \\
HV-0 (Fig.~\ref{src_vs_emc_raw_force0}) &~1.17       & --                & 0.0856$\pm$0.004 \\
HV-D                                    &~1.14       & -0.0041$\pm$0.010 & 0.0869$\pm$0.005 \\
LD (Fig.~\ref{src_vs_emc_raw})          &~0.68 (0.83)& -0.0168$\pm$0.035 & 0.0537$\pm$0.007 \\
LD-0 (Fig.~\ref{src_vs_emc_raw_force0}) &~0.61 (0.73)& --                & 0.0505$\pm$0.003 \\ 
LD-D                                    &~0.60 (0.73)& -0.0013$\pm$0.010 & 0.0508$\pm$0.003 \\

\hline

\end{tabular}
\label{fit_table}
\end{center}
\end{table}

The relevant results from the fits are summarized in Table~\ref{fit_table}.
As mentioned in the discussion of the EMC effect data, the analyses done
for JLab E03013 and SLAC E139 used different isoscalar and Coulomb distortion
corrections.  We have repeated the above comparisons of the EMC and SRC
measurements after estimating the impact of these differences and
while the numerical results change slightly (by $\approx$10\% of the
uncertainty), they do not affect the trends or the conclusions.

%% file: results.tex
\section{Potential impact of the connection} \label{results}

The close connection between the measurements of the EMC effect and the
relative contribution from short-range configurations in nuclei suggests that
the modification of the nuclear quark distributions may be related to these
short-range structures.  However, as seen in the previous section, the
connection can be made by both the HV and LD descriptions.  Future measurements
should allow us to better differentiate between these, but
at the moment, we cannot make a definitive conclusion as to the exact nature
of this connection.
In addition to helping to elucidate the origin of the EMC effect, a
better understanding of this correlation will also impact other attempts
to understand nuclear effects based on this relation.

A key aspect of the initial analysis comparing the EMC effect and
SRCs~\cite{weinstein2010rt} was the extrapolation of the EMC effect to the
free nucleon, which allows the extraction of the nuclear effects in the
deuteron.  The authors of Ref.~\cite{weinstein2010rt} use the fit to extract the IMC (in-medium correction) effect,
defined as $\frac{\sigma_A/A}{(\sigma_p +\sigma_n)/2}$, by taking the EMC
slope based on the ratio to the deuteron and adding the slope associated 
with the IMC for the deuteron, $\sigma_d/(\sigma_p+\sigma_n)$, given
by the extrapolation of the EMC/SRC linear correlation.  Given the IMC for the deuteron, they extract the sum of free
proton and neutron structure functions and, subsequently, $F_{2n}(x)$.  They obtain an IMC slope for the deuteron of 0.079$\pm$0.006
where, as discussed above, the small error is a consequence of using the known
values for the deuteron as a constraint while neglecting the correlated
uncertainties in the measurements. The equivalent global analysis from their
later work, including the new data from Ref.~\cite{fomin2012}, yields
0.084$\pm$0.004~\cite{Hen:2012fm}. In both cases, they use a fit of
the EMC slope as a function of $a_2$ which is not quite consistent
with either our LD (local density) or HV (high virtuality)
comparisons.

We repeat this extraction to obtain the IMC slope for the deuteron, using our
fits from the previous section and taking the difference of the EMC slope
extrapolated to the free nucleon ($a_2$=$R_{2N}$=0) and that for the deuteron.
Note that this is equivalent to the intercept parameter, $b$, of the fits,
and taking $dR_{IMC}(D)=b$ accounts for the correlated errors in the EMC
slopes for the deuteron and free nucleon. Similarly, one can obtain the IMC
slope for $A>2$ via $dR_{IMC}(A) = dR_{EMC}(A) + dR_{IMC}(D)$.

Our HV fit yields slopes that are close to those from the initial
analysis of~\cite{weinstein2010rt} when we apply a deuteron constraint. The unconstrained linear fit
yields a somewhat larger slope, while the LD fits all yield a smaller IMC
slope for the deuteron, suggesting smaller nuclear effects. A
reanalysis~\cite{Hen:2012fm} of the deuteron IMC effect with different data
sets found its value varied from 0.079 to 0.106, with the largest difference
associated with the use of $R_{2N}$ rather than $a_2$ from the SRC
measurements.  In the same work, the value for the IMC effect is always larger
than our results based on local density picture because they assume that only
the high-momentum nucleons associated with the SRCs contribute to the EMC
effect, while low-momentum short-distance pairs are included in our local
density analysis through the factor $N_{tot}/N_{iso}$.

The use of the SRC observables to extrapolate measurements
of the EMC effect to the free nucleon generates a large range of potential
results, with IMC slopes for the deuteron from 0.059 to 0.104, even under the
assumption that the correlation is perfectly linear all the way to $A=2$.
In addition, there is still a significant uncertainty associated with
the size of the c.m. motion correction, which modifies the extracted
values of $R_{2N}$, changing the IMC slopes for the LD extractions. This
range can be significantly narrowed if one can determine whether the
underlying connection is related to the density or the virtuality associated
with the short-distance configurations.  With further studies, this may be
possible.  If so, the nuclear effects as extrapolated from measurements can be
compared with direct calculations of the nuclear effects in the deuteron. A
recent study of the model dependence of nuclear effects in the
deuteron~\cite{Arrington:2011qt}, based on convolution calculations and
off-shell effects, produced a range of results for the neutron structure
function.  For on-shell extractions it is relatively narrow, and a direct
comparison to the IMC for the deuteron based on extrapolation from heavier
nuclei can provide a constraint on off-shell effects.

However, one must be careful in using this approach to obtain the free neutron
structure function, especially at large $x$ values.  As discussed in
Ref.~\cite{Arrington:2011qt}, extrapolations of the EMC effect to the deuteron
neglect Fermi motion, which is the dominant effect at $x>0.6$ and is sensitive
to the difference between proton and neutron structure functions at smaller
$x$ values.  Fermi motion has a significant impact and an important $Q^2$
dependence in this high-$x$ region~\cite{Arrington:2008zh, Accardi:2009br},
neither of which is accounted for in this kind of approach, limiting the reliability
of such extrapolations.  The off-shell effects as determined from the 
extrapolation of the EMC effect in Ref.~\cite{Hen:2011rt} are perfectly
consistent with the range of off-shell models included in more recent
analyses~\cite{Accardi:2011fa, Arrington:2011qt} which examine the 
model-dependence of the extraction of neutron structure functions, and the
large-$x$ deviation between the IMC-based extraction~\cite{Hen:2011rt} and the 
results of the microscopic deuteron calculations shown in that work related to
the neglect of Fermi motion in the IMC result.  Thus, it is necessary to
improve our understanding of the connection between the EMC effect and the
presence of SRCs, to better constrain the extrapolation, and to explicitly
account for both, the effects of Fermi motion and additional nuclear effects,
as done in Ref.~\cite{kulagin2006}, when going to large $x$ values.

Finally, we note that the connection between the EMC effect and SRCs suggests
a mechanism by which the structure function could have an isospin dependence
that is not included in most models.  In $^3$He, the singly-occurring neutron
is more likely to be at high momentum~\cite{arrington11}, as the SRCs are
dominantly $np$ pairs and the neutron must balance the high-momentum tail 
of both protons. The neutron also has a larger average local density: the
two-body densities from the GFMC calculations~\cite{Pieper:2001mp} show that
the two $np$ pairs have a significantly larger contribution for separations
below 1~fm than the $pp$ pair.  For both the HV and LD explanations of the
correlation between SRCs and the EMC effect, this implies a larger EMC effect
for the neutron in $^3$He, and for the proton in $^3$H.

An isospin dependence in the EMC effect for the $A$=3 nuclei would
yield an additional correction to
the neutron structure function extracted from DIS on $^3$He and
$^3$H~\cite{afnan03, E1210103}. It was shown in Ref.~\cite{afnan03} that the
difference between the nuclear effects in $^3$He and $^3$H, defined as
$\sigma_A / (Z\times\sigma_p+N\times\sigma_n)$, is extremely small, typically
less than 1\% with a spread of $\sim$1\% when varying the nuclear structure,
nucleon pdfs, and other aspects of the calculation.  Their analysis takes into
account the difference between the proton and neutron distributions in the
convolution, but not the possibility of isospin dependence in effects beyond
the convolution.   While it is unlikely to be a very large effect, given the
EMC effect for $^3$He, it may not be a negligible effect in such measurements.

Similarly, the EMC effect in heavy non-isoscalar nuclei may also have a small
isospin-dependent component.  Such an effect is generally not included in
models of the EMC effect, and would have to be accounted for in heavy nuclei
or asymmetric nuclear matter~\cite{mcgauley11}.

%% file: summary.tex
\section{Summary and conclusions}

We examined the $A$ dependence of both the EMC effect and presence of
short-range correlations in nuclei and find that the traditional models
of a simple density or $A$ dependence fail with the inclusion
of the new data on light nuclei.  Both observables show similar behavior,
suggesting a common origin.  We examined the correlation between the two
observables under two different assumptions for the underlying physics.  In the
first, we assume that the EMC effect is driven by the presence of high-momentum
nucleons in the nucleus, which is directly extracted in the inclusive 
measurements at $x>1$.  In the second, we assume that the EMC effect scales
with the average local density, and thus correlates with the number of
SRCs extracted from the $x>1$ measurements.  We find that under both
assumptions, the data are consistent with a linear correlation between the
two effects, with the local density comparison yielding a smaller $\chi^2_{\nu}$
value.  

These results support the local density explanation proposed in
Ref.~\cite{seely09}, but are still consistent with the explanation in terms of
high virtuality~\cite{weinstein2010rt}.  In the end, a more definitive
determination of the underlying physics will require further data. A large
step in this direction will be taken at JLab after the 12 GeV upgrade.  A
large set of nuclear targets, including several light nuclei with
significant cluster structure, will be used to make high precision
measurements of the EMC effect~\cite{E1210008} as well as
SRCs~\cite{E1206105}, which will further illuminate the nature of the
relationship between the two. The results from these two experiments,
combined with heavy target EMC slopes from Jlab E03-103 
will more than double the sensitivity of the linear correlation tests.

In addition, measurements probing the
modification of nucleon form factors~\cite{Paolone:2010qc,E1211002}
and structure functions~\cite{Klimenko:2005zz,E1211107} as a function of
virtuality are planned that will cover a large range of initial momentum,
allowing for direct comparison to models of the nuclear effects.